\documentclass[12pt,preprint]{emulateapj}

\newcommand{\myemail}{spd3@st-and.ac.uk}

\shorttitle{The MGC: Luminosity functions of bulges and discs}
\shortauthors{Driver et al.}

\begin{document}


\title{The Millennium Galaxy Catalogue: The luminosity functions of
  bulges and discs and their implied stellar mass densities}


\author{Simon~P.~Driver and Paul~D.~Allen}
\affil{SUPA\altaffilmark{1}, School of Physics and Astronomy, University of St Andrews, North Haugh, St Andrews, Fife, KY16 9SS, UK}
\email{\myemail}


\author{Jochen~Liske}
\affil{European Southern Observatory, Karl-Schwarzschild-Str.\ 2, 85748 Garching bei M{\" u}nchen, Germany}
\email{jliske@eso.org}

\author{Alister~W.~Graham}
\affil{Centre for Astrophysics and Supercomputing, Swinburne University of Technology, Victoria 3122, Australia}
\email{agraham@astro.swin.edu.au}

\altaffiltext{1}{Scottish Universities Physics Alliance (SUPA)}



\begin{abstract}
We derive the luminosity functions of elliptical galaxies, galaxy
bulges, galaxy pseudo-bulges and galaxy discs from our structural
catalogue of 10,095 galaxies. In addition we compute their associated
luminosity and stellar mass densities. We show that spheroidal systems
(elliptical galaxies and the bulges of disc galaxies) exhibit a strong
color bimodality indicating two distinct types of spheroid which are
separated by a core color of $(u-r) \sim 2$ mag. We argue that the
similarity of the red elliptical and the red bulge luminosity
functions supports our previous arguments that they share a common
origin and surprisingly find that the same follows for the blue
ellipticals and blue bulges, the latter of which we refer to as
pseudo-bulges. In terms of the stellar mass budget we find that
$58\pm6$ per cent is currently in the form of discs, $39\pm6$ per cent
in the form of red spheroids ($13\pm4$ per cent ellipticals, $26\pm4$
per cent bulges) and the remainder is in the form of blue spheroidal
systems ($\sim 1.5$ per cent blue ellipticals and $\sim 1.5$ per cent
pseudo-bulges). In terms of galaxy formation we argue that our data on
galaxy components strongly supports the notion of a two-stage
formation process (spheroid first, disc later) but with the additional
complexity of secular evolution occurring in quiescent discs giving
rise to two distinct bulge types: genuine 'classical' bulges and
pseudo-bulges. We therefore advocate that there are three significant
structures underpinning galaxy evolution: classical spheroids (old);
pseudo-bulges (young) and discs (intermediate). The luminous nearby
galaxy population is a mixture of these three structural types. The
nature of the blue elliptical galaxies remains unclear but one
possibility is that these constitute recently collapsed structures
supporting the notion of mass-dependent spheroid formation with
redshift.
\end{abstract}

\keywords{galaxies: spiral - galaxies: structure - galaxies:
photometry - galaxies: fundamental parameters - ISM: dust, extinction}

\section{Introduction}
In a recent paper (\citealp{mgc06}) we demonstrated that galaxy
bimodality is not just evident in color (see \citealp{strateva};
\citealp{baldry}) but also in the joint color-structure plane (see
also \citealp{ball}; \citealp{park}; \citealp{choi};
\citealp{conselice}). In that work we used SDSS photometry and single
S\'ersic (1963; \citealp{sersic}) profile fits to investigate the
distribution of 10,095 relatively nearby luminous ($M_B < -17$ mag)
galaxies in the color-S\'ersic index plane. The red peak is comprised
of highly concentrated, high-S\'ersic index systems while the blue
peak contains more diffuse, low-S\'ersic index systems. This is
important as any movement from the blue peak to the red peak will
require modifying the orbits, angular momentum and energy of the
entire stellar population. A simple inert process (e.g., exhaustion of
the gas supply, stripping, etc.) could not achieve this, although a
violent major merger event could (e.g., \citealp{lars}). Perhaps more
importantly, when the population was segregated by Hubble type we
found that while the early-type galaxies (E/S0s, i.e., bulge
dominated) lay almost exclusively in the red-concentrated peak, and
the late-type spirals (Sd/Irr, i.e., disc dominated) in the
blue-diffuse peak, the mid-type spirals (Sabc, i.e., bulge plus disc
systems) straddled both peaks with no obvious sign of bimodality. We
inferred from this that galaxy bimodality arises because of the two
component nature of galaxies and that spheroidal structures (i.e.,
ellipticals and bulges) will lie exclusively in the red-compact peak
and discs in the blue-diffuse peak. As classical bulges lie within
thin rotating disc systems this argues for early spheroid formation
(via rapid merging or collapse) followed by a more quiescent phase in
which the extended disc can form. To explore this further we have
performed bulge-disc decomposition of all 10,095 galaxies in the
Millennium Galaxy Catalogue (see \citealp{mgc08}) by fitting
two-component S\'ersic-bulge plus exponential-disc models using GIM2D
(\citealp{gim2d}). In this Letter we report the luminosity functions
derived for various component samples (e.g.\ ellipticals, bulges,
discs) and tabulate the associated luminosity and stellar mass
densities for each component class.

Throughout this paper we assume a $\Lambda$CDM cosmology with
$\Omega_{m}=0.3$, $\Omega_{\Lambda}=0.7$, and adopt $h=H_{0}/$(100 km
s$^{-1}$ Mpc$^{-1}$) for ease of comparison with other results.

\section{MGC Component Luminosity Functions}
The Millennium Galaxy Catalogue (MGC) is a deep ($\mu_{\mbox{\tiny \sc
lim}}=26$~mag~arcsec$^{-2}$), wide area ($37.5$~deg$^{2}$),
$B$-band imaging and redshift survey covering a $0.6$~deg wide strip
along the equatorial sky from 10h to 14h 50$^{\prime}$. The MGC
contains 10,095 galaxies down to $B_{\mbox{\tiny \sc mgc}}=20$~mag, of
which 9,696 have redshift information. Full details of the imaging
survey can be found in Liske et al.\ (2003), with the spectroscopic
follow-up described by Driver et al.\ (2005; hereafter D05). In Allen
et al.\ (2006) the bulge-disc decompositions were reported for all
10,095 $B_{\mbox{\tiny \sc mgc}} < 20$~mag galaxies, and a final
structural catalogue produced, consisting of bulge-only
(S\'ersic profiles), disc-only (S\'ersic or exponential
profiles) and bulge plus disc systems (S\'ersic plus exponential
profiles).

Luminosity functions are computed using the standard step-wise maximum
likelihood (SWML) estimator originally described by Efstathiou, Ellis
\& Peterson (1988), with samples divided into bins of absolute
magnitude. The MGC spectroscopic sample has a nominal Kron magnitude
limit of $B_{\mbox{\tiny \sc mgc}}=20$~mag. However, here we will use
the GIM2D total magnitudes (derived by integrating the light profiles
to infinity) so that there is no longer a single limit that applies to
the sample as a whole. To accommodate for this, each galaxy now has a
unique magnitude limit, defined by:
$B_{\mbox{lim}} = 20 + B_{\mbox{\tiny \sc
mgc}}(\mbox{S\'ersic}) - B_{\mbox{\tiny \sc mgc}}(\mbox{Kron})$.
Following D05, we restrict our sample to galaxies in the redshift
range $0.013 < z < 0.18$ and within carefully defined size and surface
brightness boundaries (see D05 and Liske et al.\ 2006 for full details).

\subsection{k+e corrections}
In D05 individual k-corrections were derived for each galaxy by
comparing the total galaxy broad-band colors ($uBgriz$) to the 27
spectral templates given in Poggianti (1998) and identifying the best
fitting spectrum. Having now separated the MGC galaxies into bulges
and discs (see \citealp{mgc08}) the global k-correction is not
necessarily valid. However, approximately $50$ per cent of our sample
are best fit by one component profiles (i.e., bulge-only or disc-only
galaxies) and for these systems we adopt the k-corrections as
previously derived in D05. For the remaining, two-component systems we
consider our component colors too coarse to be used to derive robust
individual k-corrections (as our decompositions are done in a single
filter only). For the case of blue bulges and blue discs the global
k-correction is likely to be appropriate for both (assumming that the
blue bulge has formed from the disc). In the case of discs surrounding
classical red bulges we note that for low-$B/T$ systems the
k-correction is likely to be appropriate for the discs but not the
bulges. In the case of high-$B/T$ systems we note that Peletier \&
Bacells (1996) report that such discs are typically redder. We
therefore consider it appropriate to continue to adopt the global
k-correction for both our discs and blue bulges. For the red bulges,
however, we adopt the spectral template most frequently adopted by our
single component red ellipticals (a Sa 15 Gyr spectrum). This spectrum
can be represented by a fourth order polynomial valid over the
redshift range $0 < z < 0.18$ only:
\begin{equation} 
k(z)=3.86z+12.13z^{2}-50.14z^{3}.
\end{equation} 
We note that if we follow a similar procedure for the discs and
blue bulges the implied characteristic turnover luminosities (shown in
Table~1) are systematically reduced by $\sim 0.1$~mag and the implied
stellar masses reduced by $\sim 14$ per cent.

To model evolution we assume pure luminosity evolution of the
following form:
\begin{equation}
L_{z=0}=L(1+z)^{-\beta},
\end{equation}
where $\beta$ is set to $0.75$ for the global luminosity function
\citep[see][]{mgc05}, $1$ for blue components (discs, blue bulges,
blue ellipticals), and $0.5$ for red components (red bulges and red
ellipticals). These values are based on the recent results reported in
Zucca et al.\ (2006) for red and blue systems (their type 1 and 4
respectively). We do not model number evolution as the redshift range
is small and our merger rate estimates, based on dynamically close
pairs within the MGC, are low (see \citealp{mgc10} and
\citealp{mgc13}).

\subsection{Luminosity functions}
Fig.~1 shows the luminosity distributions and Schechter function fits
for our full galaxy sample (upper left), ellipticals only (upper
right; i.e.\ objects with $B/T=1$ after logical filtering, see Fig.~13
of Allen et al.\ 2006), discs (lower left) and bulges (lower
right). Note that the Schechter function in the upper left differs
from that shown in D05 because the magnitudes are now based on
S\'ersic profiles integrated to infinity rather than Kron
magnitudes. This difference is significant, resulting in a brighter
$M^*$ value by about $0.1$~mag but a comparable faint-end slope,
$\alpha$. Analysis of independent repeat observations of $\sim 700$
galaxies suggests that our decompositions are valid to good accuracy
($\Delta M_{\mbox{\tiny \sc bulge}} = \pm 0.1$~mag and $\Delta
M_{\mbox{\tiny \sc disc}} = \pm 0.15$~mag, see Allen et al.\ 2006) for
componentss with luminosities with $M_B < -17$~mag. Below this limit
our decompositions become increasingly less reliable and these data
are shown with open symbols. The most striking result from Fig.~1 is
the rapidly rising faint-end slope for the elliptical population. This
was noted previously in Driver et al.\ (2006) and was shown to be due
to contamination of the classical elliptical sample by low luminosity
blue spheroids (see also \citealp{ellis05}).


In Fig.~2 we show the color-structure plane defined by SDSS core
($u-r$) PSF color versus component S\'ersic index for the ellipticals
(upper left) and galaxy bulges (upper right). The bimodality of the
ellipticals is striking, with a blue and red population being
apparent. The blue sample defines what we label blue ellipticals which
were identified in Ellis et al.\ (2005) and quantified in Driver et
al.\ (2006).  A cut at $(u-r)=2$~mag provides a clear division. The
lower panels of Fig.~2 show the luminosity distributions and Schechter
function fits for the blue and red samples. We see that red
ellipticals and blue ellipticals follow markedly different trends and
it is indeed the blue ellipticals which are responsible for the
apparent upturn in the total elliptical galaxy luminosity function at
very faint absolute magnitudes. We note that when the bulges are
divided in the same manner, the blue bulge luminosity function is very
similar to that of the ellipticals, possibly indicating some common
origin.

It is tempting to associate the blue bulges with pseudo-bulges (see
Kormendy \& Kennicutt 2004), which are believed to arise from inner
disc instabilities giving rise to a 'swelling' of the disc in the
central region. As many of our blue bulges have $M_B > -17$~mag, where
our bulge-disc decompositions become unreliable, we cannot
unambiguously confirm this population as pseudo-bulges but for the
moment adopt this nomenclature for ease of discussion. The blue
ellipticals remain somewhat intriguing and appear to define a new
class of object as previously noted by Ellis et al.\ (2005) and Driver
et al.\ (2006; see also \citealp{vdds} who identify them as a rapidly
fading population).  We are currently exploring these systems further
(\citealp{ellis07}) and for the moment simply flag them as
interesting.  From their distinct luminosity function it is clear they
are predominantly low luminosity systems and could potentially
represent the local counterparts to the luminous blue compact galaxies
studied by Guzman et al.\ (1997) and Phillips et al.\ (1997).

Fig.~3 shows the final component luminosity functions with the red
ellipticals and red bulges combined into a single red spheroid group
and the blue ellipticals and blue bulges grouped together into a
single blue spheroid class. The justification for this is the
similarity in the shapes and ranges of the luminosity distributions
from Fig.~2. 

\subsection{Luminosity densities and stellar-mass densities}
Table~1 shows the Schechter function values for the luminosity
function fits shown in Figs.~1, 2 \& 3, along with their associated
$b_J$-band luminosity, $j_{b_J}$, and stellar mass densities,
$\rho_{\mathcal{M}}$. These are derived using the following
expressions:
\begin{equation}
j_{b_{J}}=\phi^{*}10^{-0.4(M^*_{B}-M_{\odot})}\Gamma(\alpha+2)
\end{equation}
and
\begin{equation}
\rho_{\mathcal{M}}=\sum^{N}_{i}\,(\phi_{i}/N_{i})10^{(1.93(g-r)_i-0.79)}10^{-0.4(M_{B,i}-M_{\odot})}.
\end{equation}
The latter expression is first shown in Driver et al.\ (2006) and is
based on the color to mass-to-light ratios given by Bell \& de Jong
(2001) which assume a Salpeter-'lite' IMF. Note that the $(g-r)$
colour for each galaxy was obtained by matching the MGC to the Sloan
Digital Sky Survey first data release (Abazajian et al. 2003).

\section{Discussion}
From Table~1 we see that $58\pm6$ per cent of the stellar mass is in
the form of galaxy discs, $13\pm4$ per cent in red elliptical galaxies
and $26\pm4$ per cent in classical red bulges. Previously it has been
reported (Bell et al.\ 2003; Baldry et al.\ 2004; Driver et al.\ 2006)
that $54-60$ per cent of the stellar mass lies in the red peak
dominated by early-type galaxies. As we have now separated the
early-type galaxies into bulges and discs one expects that the
spheroid (elliptical+bulge) stellar mass density should be lower than
the early-type stellar mass density, which of course includes the
associated disc components of the lenticular galaxies. Examining the
color-structure plane (Fig.~2 upper panels) for ellipticals and bulges
we see that the red populations (of each type) lie in the same
location. This is indicative of a shared origin for red ellipticals
and classical red bulges. When combined the red spheroids account for
$39\pm6$ per cent of the stellar mass density. Hence the bulk of the
stellar mass ($96$ per cent) exists in the two classical structures
originally defined by de Vaucouleurs (1959).

The remaining $3$ per cent lie in the form of blue ellipticals ($\sim
1.5$ per cent) and blue bulges ($\sim 1.5$ per cent). These latter two
populations can therefore be considered minor from a cosmological
perspective. Furthermore, as the blue bulges are likely to represent
either difficulties in the decomposition (e.g., bars) or pseudo-bulges
(disc swelling), their stellar mass could arguably be added, in either
case, to that of the galaxy discs (i.e., $\sim61$ per cent). The
nature of the blue ellipticals (previously dubbed blue spheroids)
remains uncertain but they appear to constitute a very small fraction
of the stellar mass budget, although we must note the near divergent
faint-end slopes. The primary conclusion then is that the stellar mass
is mainly divided between two distinct structures: blue 2D discs and
red 3D spheroids.

Finally we note that the results presented in this Letter are based on
$B$-band data and therefore susceptible to dust attenuation (see
\citealp{shao}). Bell \& de Jong (2001) argue that the effect on an
individual galaxy's stellar mass is less than one might expect (see
their fig.~1) because the observed decrease in total luminosity is
offset by the increased stellar mass-to-light ratio inferred from the
redenned colors, therefore yielding comparable final stellar
masses. We explore this in detail in Driver et al.\ (2007) and note
that while the masses are modified somewhat the final stellar mass
breakdown is not dramatically altered.

\acknowledgments
The Millennium Galaxy Catalogue consists of imaging data from the
Isaac Newton Telescope and spectroscopic data from the Anglo
Australian Telescope, the ANU 2.3m, the ESO New Technology Telescope,
the Telescopio Nazionale Galileo, and the Gemini Telescope. The survey
has been supported through grants from the Particle Physics and
Astronomy Research Council (UK) and the Australian Research Council
(AUS). The data and data products are publicly available from
http://www.eso.org/$\sim$jliske/mgc/ or on request from J.~Liske or
S.P.~Driver.

{}

\begin{deluxetable}{lcccccc}
\setlength{\tabcolsep}{0.025in}
\tabletypesize{\scriptsize}
\tablecolumns{7}
\tablehead{\colhead{Sample} & \colhead{$M^{*}_{B} - 5\log\,h$} & \colhead{$\alpha$} & \colhead{$\phi^{*}$}  & \colhead{$j_{b_{J}}$
} & \colhead{$\rho_{\mathcal{M}}$} & \colhead{$N$} \\
& \colhead{(mag)} & & \colhead{($10^{-2} \, h^3$ Mpc$^{-3} \, (0.5
    $ mag$)^{-1}$)} & \colhead{ ($10^8 \, h \, L_{\odot}$ Mpc$^{-3}$)} & 
\colhead{  ($10^8 \, h \, M_{\odot}$ Mpc$^{-3}$)} & }
\tablecaption{Schechter function parameters, luminosity densities, and stellar 
mass densities of the various component sub-divisions discussed in
this paper.}
\startdata
All                         & $-19.84\pm0.02$ & $-1.15\pm0.01$ & $1.72\pm0.05$ & $2.65\pm0.13$   & $6.2\pm0.3 $ & $7786$\\
All Discs                   & $-19.44\pm0.04$ & $-1.15\pm0.03$ & $1.74\pm0.09$ & $1.85\pm0.20$   & $3.6\pm0.4 $ & $6024$\\
Bulges                      & $-19.23\pm0.07$ & $-1.00\pm0.08$ & $0.64\pm0.05$ & $0.50\pm0.11$   & $1.7\pm0.4 $ & $1431$\\
Ellipticals                 & $-19.36\pm0.12$ & $-0.91\pm0.11$ & $0.38\pm0.05$ & $0.32\pm0.11$   & $0.9\pm0.3$ & $ 835$\\
Blue Ellipticals            & $-19.89\pm0.50$ & $-1.88\pm0.22$ & $0.02\pm0.01$ & $0.11\pm\infty$\tablenotemark{a} & $0.1\pm\infty$\tablenotemark{a} & $ 229$\\
Red Ellipticals             & $-19.02\pm0.11$ & $-0.26\pm0.13$ & $0.36\pm0.02$ & $0.21\pm0.03$   & $0.8\pm0.2$ & $ 606$\\
Blue Bulges                 & $-19.87\pm0.50$ & $-2.08\pm0.21$ & $0.02\pm0.01$ & $0.92\pm\infty$\tablenotemark{a} & $0.1\pm\infty$\tablenotemark{a} & $ 249$\\
Red Bulges                  & $-19.11\pm0.07$ & $-0.75\pm0.08$ & $0.65\pm0.05$ & $0.42\pm0.07$   & $1.6\pm0.3 $ & $1182$\\
Ellipticals +  Bulges       & $-19.34\pm0.07$ & $-1.01\pm0.06$ & $0.96\pm0.08$ & $0.84\pm0.16$   & $2.7\pm0.5 $ & $2266$\\
Red (Ellipticals + Bulges)  & $-19.16\pm0.07$ & $-0.67\pm0.07$ & $0.97\pm0.06$ & $0.64\pm0.08$   & $2.4\pm0.3 $ & $1788$\\
Blue (Ellipticals + Bulges) & $-19.89\pm0.32$ & $-1.97\pm0.14$ & $0.04\pm0.02$ & $0.20\pm\infty$\tablenotemark{a} & $0.3\pm\infty$\tablenotemark{a} & $ 478$\\
\enddata

\tablenotetext{a}{These luminosity functions are potentially divergent
within the specified errors, giving rise to extreme luminosity and
stellar mass densities. As we do not know whether these distributions
continue to diverge we infer the luminosity and stellar mass densities
required for a fully self-consistent table, i.e., the stellar mass of
blue spheroids is derived by subtracting the red ellipticals' stellar
mass from the total ellipticals' stellar mass, etc.}

\end{deluxetable}

\clearpage

\begin{figure}
\plotone{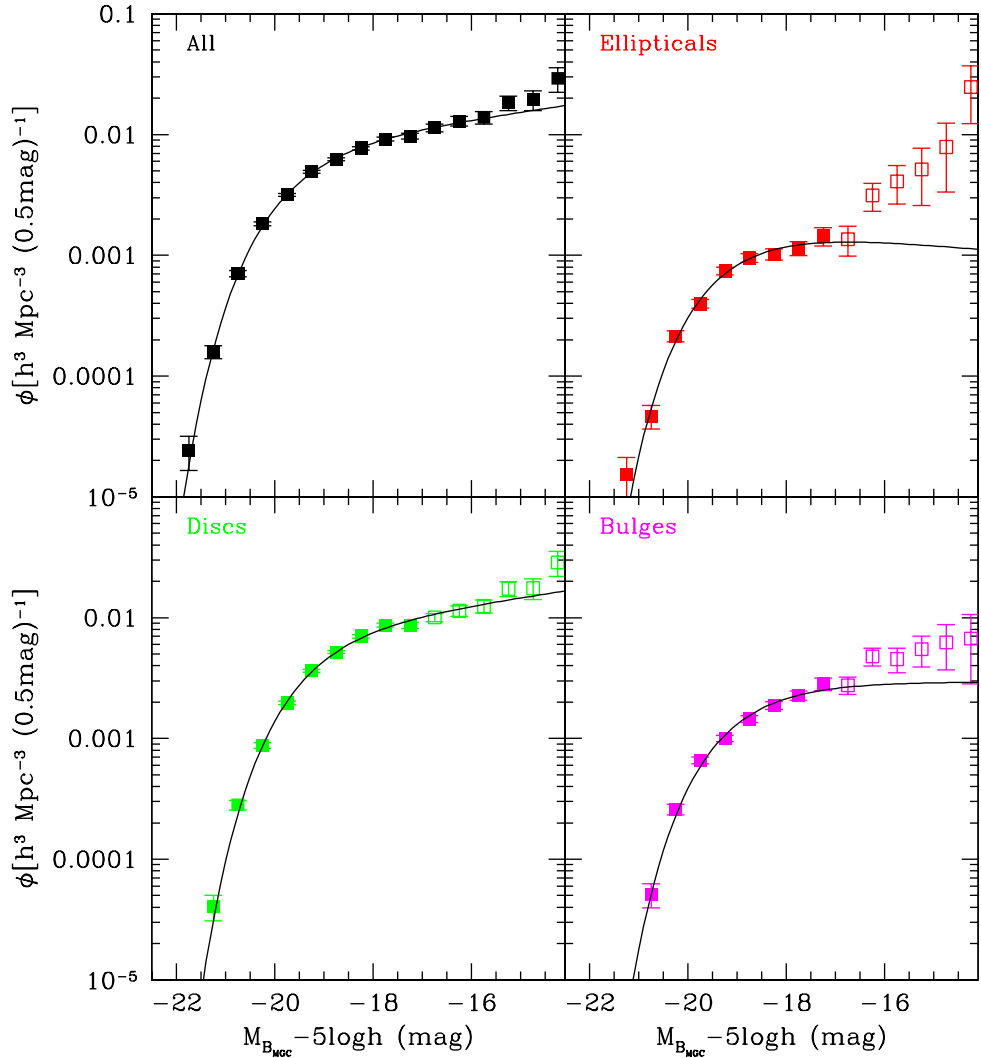}
\caption{$B$-band luminosity functions derived using our variable
limit SWML method for the global sample (top left), discs (bottom
left), ellipticals (top right), and bulges (bottom right). In each
case the derived data (squares), and fitted Schechter functions (solid
lines) are shown. The Schechter fit for the global sample uses all
data points but for the component luminosity functions (discs,
ellipticals, and bulges) we only use data points with $M_{B}<-17$~mag
(see Section 2.3).}
\end{figure}

\clearpage

\begin{figure}
\plotone{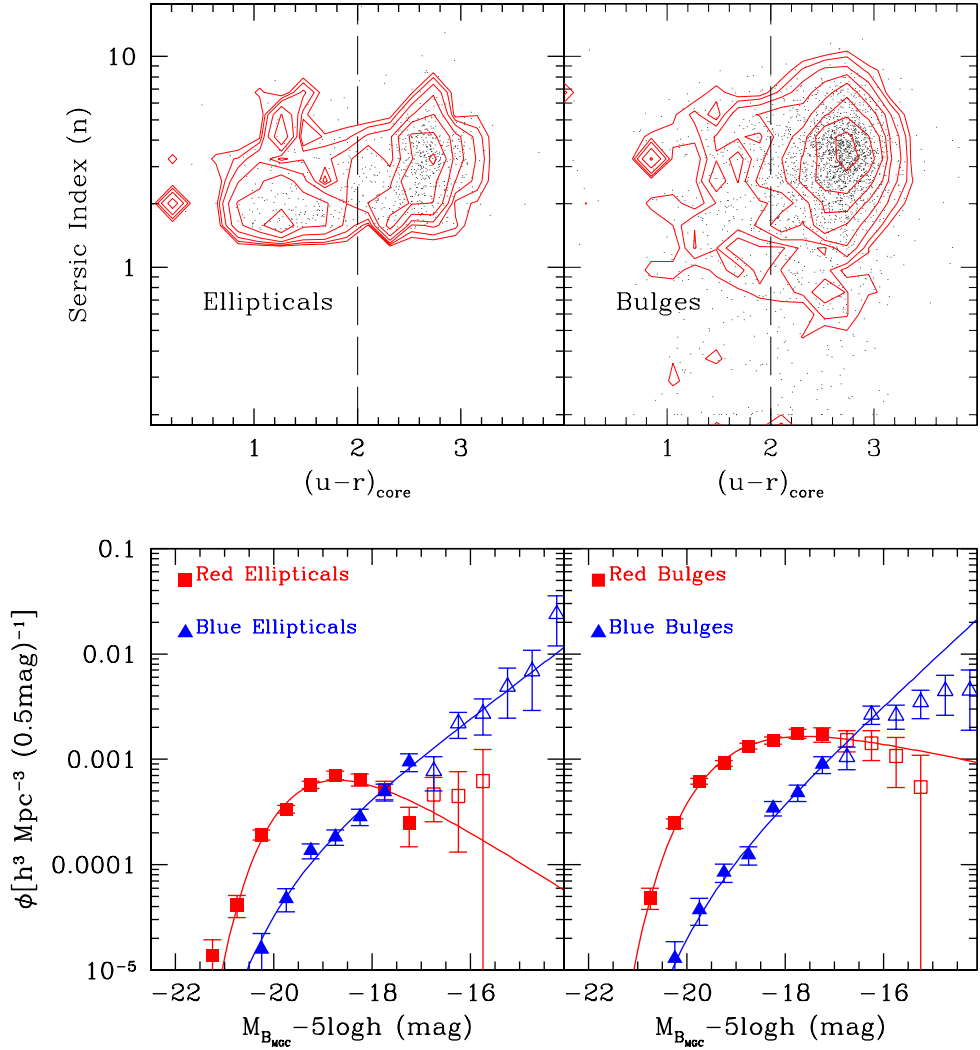}
\caption{The top row shows the $(u-r)-n$ distribution (dots) for
ellipticals (left), and bulges (right) with $M_B < -17$~mag. The
contours show the volume-corrected luminosity density for these
objects. The natural division between the red and blue populations
appears to occur at $(u-r)=2$~mag. The bottom row shows the $B$-band
luminosity functions for ellipticals (left), and bulges (right), split
into red, $(u-r)>2$~mag, components where data points are shown by
squares, and blue, $(u-r)\le2$ mag, components where data points are
shown by triangles.}
\end{figure}

\clearpage

\begin{figure}
\plotone{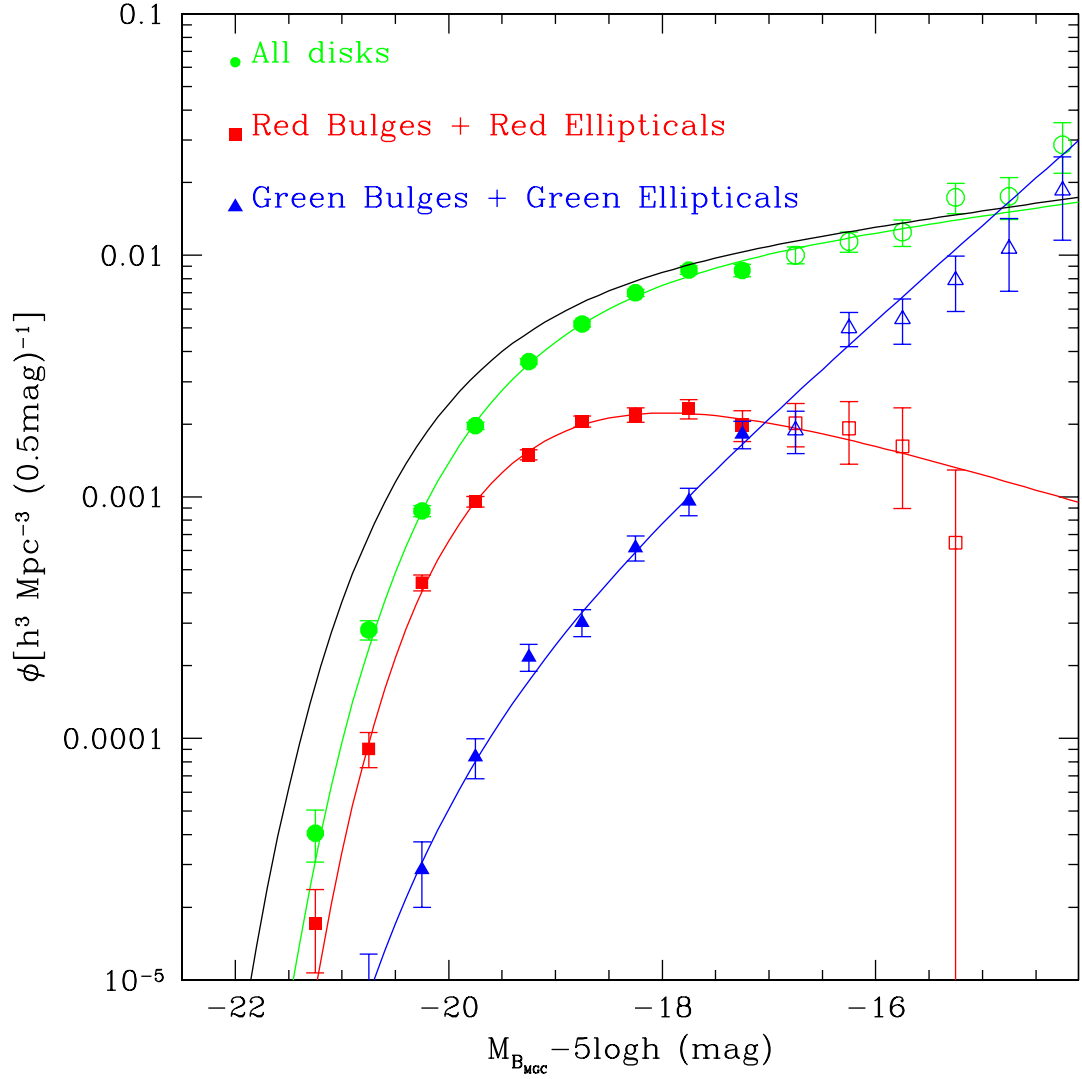}
\caption{Derived data and Schechter fits for $B$-band luminosity
functions for our three final structural components: discs (circles),
red bulges and ellipticals (squares), and blue (pseudo-)bulges and
blue ellipticals (triangles). The Schechter fits only use data points
with $M_{B}<-17$~mag. The thicker black line shows the global $B$-band
luminosity function.}
\end{figure}

\clearpage

\end{document}